\documentclass[final,authoryear,5p,times,twocolumn]{elsarticle}

\def\astrobj#1{#1}

\usepackage{graphicx}

\usepackage{amssymb}
\usepackage{amsmath}
\usepackage{multirow}

\usepackage{ccaption}

\journal{New Astronomy}

\begin{document}

\begin{frontmatter}



\title{Computational advances in gravitational microlensing: a comparison of
CPU, GPU, and parallel, large data codes\tnoteref{label1}}
\tnotetext[label1]{Research undertaken as part of the Commonwealth Cosmology Initiative (CCI: www.thecci.org), an international collaboration supported by the Australian Research Council}


\author[swin]{N.F.~Bate}
\ead{nbate@astro.swin.edu.au}

\author[swin]{C.J.~Fluke}
\ead{cfluke@astro.swin.edu.au}

\author[swin]{B.R.~Barsdell}

\author[usyd]{H.~Garsden}

\author[usyd]{G.F.~Lewis}

\address[swin]{Centre for Astrophysics and Supercomputing, Swinburne
  University of Technology, P.O. Box 218, Hawthorn, Victoria, 3122, Australia}

\address[usyd]{Sydney Institute for Astronomy, School of Physics, A28,
  The University of Sydney, NSW 2006, Australia} 

\begin{abstract}
To assess how future progress in gravitational microlensing computation 
at high optical depth will rely on both hardware and software solutions, 
we compare a direct 
inverse ray-shooting code implemented on a graphics processing unit (GPU) 
with both a widely-used hierarchical tree code on a single-core CPU, and 
a recent implementation of a parallel tree code suitable for a
CPU-based cluster
supercomputer.  We examine the accuracy 
of the tree codes through comparison with a direct code over a much 
wider range of parameter space than has been feasible before. We demonstrate 
that all three codes present comparable accuracy, and choice of approach 
depends on  considerations relating to the scale and nature of the microlensing
problem under investigation.
On current hardware, there is little difference in the processing speed 
of the single-core CPU tree code and the GPU direct code, however the
recent plateau in single-core CPU speeds means the existing tree code is no longer able 
to take advantage of Moore's law-like increases in processing speed. 
Instead, we anticipate a rapid increase in GPU capabilities in the next 
few years, which is advantageous to the direct code. 
We suggest that progress in other areas 
of astrophysical computation may benefit from a transition to GPUs 
through the use of ``brute force'' algorithms, rather than attempting 
to port the current best solution directly to a GPU language -- 
for certain classes of problems, the simple implementation on GPUs may 
already be no worse than an optimised single-core CPU version.
\end{abstract}


\end{frontmatter}


\section{Introduction}
\label{sct:intro}

Gravitational microlensing is the deflection of light by matter in 
the regime where multiple-imaging and high magnification events occur, but
the angular separation of images is too small to resolve.  The observable
effect is a change in brightness of a source over a period of time, with the
details of the light curve depending on the complexity of the lens
distribution and the surface brightness profile of the source.
Gravitational microlensing has been used to study high magnification events
on local scales, viz. the searches for dark matter compact objects in
the Galactic bulge and halo (see for example references in and
citations to \citealt{alcock+93}; \citealt{aubourg+93};
\citealt{udalski+93}; \citealt{calchi+02}; \citealt{wyrzykowski+09}) and extrasolar planets (e.g. \citealt{gould+92};
\citealt{bond+04}; \citealt{sumi+10}), and quasar microlensing on cosmological scales (e.g. \citealt{vanderriest+89}; \citealt{irwin+89};
\citealt{k04}; \citealt{bate+08}; \citealt{dai+10}). For recent
reviews of microlensing on all scales, see \citet{wambsganss06}, 
\citet{kochanek+07}, \citet{gould08}, \citet{mao08} and
\citet{gaudi10}.

Microlensing on local scales occurs in the low optical depth regime,
where the source is affected by at most a few microlenses along the line of
sight. Large numbers of stars must therefore be monitored in order to
observe a single microlensing event. Conversely, the microlensing optical
depth at the location of lensed quasar images is likely to be
$\sim1$. This microlensing signal is therefore more complex, as it is
the result of a large number of microlenses acting on the source
essentially all the time \citep{wambsganss06}. In the remainder of this
work, we focus on high optical depth microlensing. 

\begin{figure}
  \centering
  \scalebox{0.48}{\includegraphics{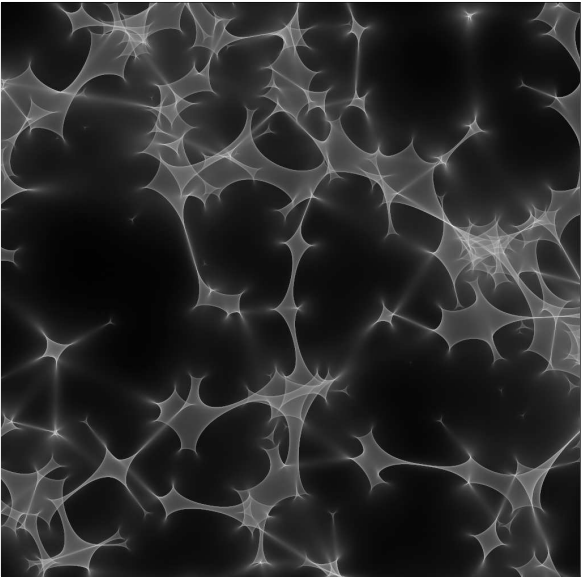}}
  \caption{A sample microlensing magnification map for a microlensing model
with the following parameters: total convergence $\sigma = 0.4$, shear $\gamma=0.0$,
$N_{\rm pix} = 1024^2$ pixels in the source plane and $N_{\rm rays} =
2000$ light rays per source pixel (on average). These parameters
correspond to $N_{*} = 558$ lenses. The source plane has a side length
of 20 Einstein radii. The greyscale covers the range from low magnification (black)
to high magnification (white). }
\label{fig:map}
\end{figure}

A key element in many microlensing calculations is the determination of the
magnification map; a typical example is shown in Figure \ref{fig:map}.  
This map provides an estimate of the point-source brightening over a finite 
region of a two-dimensional source plane.  The magnification of an extended 
source is found by convolving a source profile (intensity and geometry) with
the magnification map.  This approach has been used to probe the
structure of quasar emission regions, including accretion discs
(e.g. \citealt{k04}; \citealt{eigenbrod+08}; \citealt{floyd+09}) and
broad emission line regions (e.g. \citealt{lewis+04}; \citealt{abajas+07}).

The limitation in many microlensing calculations is the computational 
overhead in producing magnification maps with sufficient resolution and 
high (statistical) accuracy. Fitting observed light-curves obtained over long periods of time requires large maps with high pixel resolution to resolve small sources (such as quasar X-ray emission regions). Similarly, fitting multi-wavelength observations requires both large maps and high pixel resolution in order to simultaneously model emission from small optical emission regions and larger infrared emission regions. The situation is even worse when attempts are made to simultaneously model microlensing of accretion discs and the much larger broad emission line regions, a difference in scale of 2--3 orders of magnitude. Finally, many maps must be generated in order to ensure statistical independence of simulated measurements. It is only reasonable to sample any one magnification map as many times as you can place a source on it without overlapping.

In this paper we compare three techniques for computing magnification maps:
a hierarchical tree code for single-core CPUs; a parallel, large data 
tree code for use on a CPU-based cluster supercomputer; and a direct, 
inverse ray-shooting code for use on graphics processing units (GPUs).  
We consider the accuracy of the codes, to determine whether the
approximations inherent in the tree code are valid over a wider range
of parameter space than has been feasible to test before, and compare
run-times between the single-core tree code and the GPU code over a range
of astrophysically-motivated parameters.  While the ``brute force'' approach
to ray shooting has not been considered a feasible option on CPU architectures,
we show that the GPU implementation is highly competitive, and is indeed
faster than the tree code in many circumstances.   The expected increase
in GPU speeds in the years ahead bodes well for advances in gravitational
microlensing simulations. Our work hints at the potential benefit to 
other fields of astrophysical computation from the implementation of simple, 
``brute force'' algorithms as a means of accelerating the adoption 
of GPUs in astronomy for time-consuming computations. 

The remainder of this paper is set out as follows.  In Section \ref{sct:magmap},
we describe the process of generating magnification maps, and 
summarise key features of three gravitational microlensing codes which 
can be used for this task.  In Section \ref{sct:results},
we present a series of code comparisons (accuracy and speed). In
Section \ref{sct:discussion}, we discuss 
the relative merits of hierarchical and direct ray-shooting approaches, 
and suggest the classes of problems to which each is best-suited.
We present our conclusions in Section 
\ref{sct:conclusion}. 

\section{Generating magnification maps}
\label{sct:magmap}
In general, magnification maps are calculated using the (two-dimensional) 
gravitational lens equation
\begin{equation}
{\mathbf y} = {\mathbf x} - \boldsymbol{\alpha} ({\mathbf x}),
\label{eqn:lens}
\end{equation}
which relates the locations of the source, ${\mathbf y}$, and 
an image, ${\mathbf x}$,
via the deflection angle, $\boldsymbol{\alpha} ({\mathbf x})$. 
The magnification is:
\begin{equation}
\mu = 1/\det {\mathbf A}
\end{equation}
where ${\mathbf A} = \partial {\mathbf y}/\partial {\mathbf x}$ is the Jacobian
matrix of equation (\ref{eqn:lens}).

Equation (\ref{eqn:lens}) gives a one-to-one relationship between an image 
location and the corresponding source location (${\mathbf x} \rightarrow
{\mathbf y}$), but a one-to-many relationship for the alternative case of mapping
a source location to its multiple images (${\mathbf y} \rightarrow {\mathbf x}$). 
While analytic solutions exist
for a small number of simple lens models (e.g. \citealt{schneider+92}), in general 
a numerical approach is used to calculate $\mu$ over a finite region
of the source plane. This can be achieved most easily
using inverse ray-shooting (\citealt{kayser+86};
\citealt{schneider+86}; \citealt{schneider+87}).
Here, light rays are propagated backwards from the observer,
through the lens plane, where they are deflected using equation (\ref{eqn:lens}),
and then mapped to a grid of pixels in the source plane. This approach avoids
the need to determine the location of every image.  By counting the number
of light rays reaching each source plane pixel, $N_{ij}$, compared to 
the average per pixel if there was no lensing, $N_{\rm rays}$, an estimate 
of the per-pixel magnification, $\mu_{ij}$, is obtained:
\begin{equation}
\mu_{ij} = N_{ij}/N_{\rm rays}.
\end{equation}

For cosmological microlensing problems, the appropriate model for the deflection angle describes $N_{*}$ compact 
lens masses in a smooth background mass distribution, acted on 
by an external (macro-model) shear $\gamma$. This model, an 
extension of the one proposed by \citet{chang+79}, is suitable for 
considering the impact of lensing due to individual stars where the external
shear is due to the large-scale galactic mass distribution.
The lens equation in this case is:
\begin{equation}
{\mathbf y} = \left(
\begin{array}{cc}
1 - \gamma & 0 \\
0 & 1 + \gamma
\end{array}
\right) {\mathbf x} - \sigma_c {\mathbf x}
- \sum_{i=1}^{N_*} m_i \frac{({\mathbf x} - {\mathbf x}_i ) }
{\vert{\mathbf x} - {\mathbf x}_i\vert^2}
\label{eqn:sigma}
\end{equation}
where the convergence $\sigma = \sigma_c + \sigma_*$ combines contributions from smooth 
matter $\sigma_c$, and compact objects $\sigma_*$.  This is the model
for the deflection of light by a screen of mass that we will use throughout this paper.

The relationship between $\sigma_*$ and the number of lenses $N_*$ is:
\begin{equation}
N_* = \frac{\sigma_*A}{\pi\langle M \rangle }
\label{eqn:nstar}
\end{equation}
If the deflecting mass distribution were smoothed out, a circle in the
image plane would map onto an ellipse with major-to-minor axis ratio
$|1-\sigma+\gamma|/|1-\sigma-\gamma|$ in the lens plane. The area of
this `minimum' ellipse is set by the size of the image plane. In practice, the graininess of
the matter distribution can cause rays to be scattered outside the
receiving area. Similarly, rays from a long way outside the shooting
region can be scattered into the receiving square in the source
plane. For that reason, the lenses in the codes considered here are
distributed in a circle of area $A$ with a diameter larger than the long axis of
the minimum ellipse.

The number of calculations, $\Phi$,  required
to obtain a mangification map with $N_{\rm pix}$ pixels
scales as:
\begin{equation}
\Phi = N_{\rm FLOP} \times N_{\rm pix} \times N_{\rm *} 
\times N_{\rm rays},
\end{equation}
where $N_{\rm FLOP}$ is the number of floating point operations for each
deflection calculation, and $N_{\rm rays}$ is chosen to achieve a desired level
of accuracy.  It was quickly realised that the direct ray-shooting approach
was not practical on single CPUs: a scenario, $\Phi_{\rm fiducial} \sim 3 \times 10^{16}$, 
with $N_{\rm pix} = 2500^2$ pixels, $N_{\rm rays} = 500$, 
$N_{*} = 10^6$ lenses,\footnote{Relevant for the case $\sigma \sim 1$}
and $N_{\rm FLOP} = 10$ would take months to years to calculate 
\citep{wambsganss99} on hardware that was available at the time.

Wambsganss (1990; 1999) introduced a method based on a 
hierarchical tree code \citep{barnes+86}. Here, the contribution
of each lens depends on its distance from the light ray -- lenses
that are further away are grouped together into pseudo-lenses with a
higher mass. This approach reduces $N_{*}$, thus increasing the speed
of the overall calculation, but at the cost of a slight error in the magnification
map due to the approximation.  This approach has been used
to constrain both the size and temperature profile of quasar accretion
discs (e.g. \citealt{wambsganss+90}; \citealt{rauch+91}; \citealt{morgan+06}; \citealt{pooley+07}; \citealt{keeton+06}; \citealt{anguita+08}; \citealt{eigenbrod+08}; \citealt{bate+08}; \citealt{floyd+09}), dark matter fractions in the lensing
galaxies (e.g. \citealt{pooley+09}; \citealt{dai+10}), and possibly
even the orientation of the quasar with respect to the line of sight \citep{poindexter+10}.

Despite the much-improved speed compared to direct ray-shooting, 
the hierarchical method is complex to implement, and requires a
certain amount of expert knowledge. These facts make slight 
modifications or additions to the code quite challenging.  Parameters
such as the required accuracy (which determines whether to use a
particular lens cell, or its four subcells, based on the distance
between the ray and the cell's centre of mass) must be determined
experimentally by repeatedly running the code until acceptable results
are obtained. Additionally,
the generation of the tree data structure imposes a memory overhead which exceeds
typical single CPU memory sizes for large $N_*$. A parallel hierarchical tree code, suitable for running billion-lens 
calculations, has now been developed by \citet{garsden+10}.   
This approach overcomes the single-CPU memory limitation by distributing 
the computation over a number of nodes in a CPU-based cluster supercomputer.

Alternatives to the hierarchical method include the use of polygonal cells
\citep{mediavilla+06}, optimised to map areas of the image plane to the 
source plane thus reducing $N_{\rm rays}$ required for a given accuracy, and the
contouring method for point sources (\citealt{lewis+93}; \citealt{witt93}), 
which maps an infinite line plus 
$N_{*}$ closed loops from the image to source plane.  The polygonal
cell approach results in its own computational overhead in determining
the correct polygonal lattice, and suffers from the need for a
substantial number of polygons in the vicinity of caustics, which
results in lengthy computation times. The contouring method has limited
applicability to extended sources. \citet{wyithe+99} presented an
extension of the contouring method to cover extended sources,
however the additional computations required slowed the code to
the point where it could no longer compete with the simplicity of
inverse ray-shooting.

A recent alternative is the appearance of the GPU as a mathematical co-processor. 
Originally developed to enhance the rate of generating graphics for 
the computer game industry, GPUs are revolutionising scientific 
computing (\citealt{fournier+88}; \citealt{tomov+03};
\citealt{venkat+03}; \citealt{owens+05}).  Their highly parallel 
processing architecture, accessed
through flexibile programming libraries such as NVIDIA's
CUDA\footnote{{\tt http://www.nvidia.com/object/cuda\_home.html}} 
or the cross-platform OpenCL\footnote{{\tt http://www.khronos.org/opencl}} 
standard created by the Khronos Group,
is enabling speed-ups of $O(10)$ to $O(100)$ times over a broad range 
of algorithms.  Indeed, the notion of general purpose computing on 
graphics processing units (GPGPU) has motivated a reinvestigation of 
time-consuming processing tasks.  

Early adoption of GPGPU in astronomy has included $N$-body simulations (e.g.,
\citealt{belleman+08}), radio-telescope signal correlation (e.g.,
\citealt{harris+08}), the solution of Kepler's equations \citep{ford08},
radiation-transfer models \citep{jonsson+09} and
adaptive mesh refinement simulations \citep{schive+10}.

From a conceptual and implementational point of view, direct ray-shooting
is a very simple algorithm with a high degree of parallelism: the deflection 
of each light ray is independent of every other light ray, and the 
deflection of a light ray due to a single lens is independent of all 
other lenses.    The existence of such fine-grained parallelism means that
ray-shooting can be considered an ``embarassingly parallel'' algorithm --
ideal for GPGPU.

\citet{thompson+10} reported on a CUDA implementation of 
direct ray-shooting,
and demonstrated that their code could calculate a magnification map for 
the $\Phi_{\rm fiducial}$ scenario in $\sim 7$ hours using a four-GPU
NVIDIA S1070 Tesla unit.
Moreover, they were able to calculate a magnification map 
with one billion lenses ($N_{\rm pix} = 512^2$ and $N_{\rm rays} = 42$) 
in 24 hours, with a peak sustained processing rate of 1.28 Tflop/sec. 
While the CUDA implementation of direct ray-shooting gave an $O(100)$ 
processing speed-up compared to using the same method on a single CPU, a 
critical comparison omitted from \citet{thompson+10} was between 
the tree code and the GPU code.  

In this paper, we compare:
\begin{itemize} 
\item The ``industry standard'' single-core tree code by Wambsganss (1990;
1999), which we refer to as CPU-T;
\item The parallel, large-data tree code by \citet{garsden+10}, which
we refer to as CPU-P; and
\item The CUDA direct ray-shooting code by \citet{thompson+10}, which
we refer to as GPU-D.
\end{itemize}
We now describe salient points of these three approaches in more detail,
including implementation issues that are relevant for the code comparison
in Section \ref{sct:results}.

\subsection{CPU-T}
The most computationally intensive part of the ray-shooting algorithm is the calculation of the deflection of individual rays by each lens. The hierarchical tree code attempts to minimise the number of calculations necessary by grouping together lenses 
that are sufficiently distant from a given light ray. A group of lenses is treated as a single pseudo-lens with a total mass equal to the sum of the individual lens masses, located at their centre of mass.

The first step in the calculation is to divide up the lenses into a cell structure, starting with the root cell, which is the size of the lens plane. Each cell that contains more than one lens is subdivided into four cells with half the side length of the parent cell.  This process is continued iteratively until only cells containing zero or one lens remain. This calculation need only be carried out once per magnification map, and typically takes only a small fraction of the code's total run-time. 
A determination must be made for each light ray regarding which lenses need to 
be included individually and which can be grouped together into pseudo-lenses. The lens equation then becomes:
\begin{equation}
{\mathbf y} = \left(
\begin{array}{cc}
1 - \gamma & 0 \\
0 & 1 + \gamma
\end{array}
\right) {\mathbf x} - \sigma_c {\mathbf x}
- \sum_{i=1}^{N_L} m_i \frac{({\mathbf x} - {\mathbf x}_i ) }
{\vert{\mathbf x} - {\mathbf x}_i\vert^2} - \sum_{j=1}^{N_{PL}} m_j^{PL} \frac{({\mathbf x} - {\mathbf x}_j^{PL} ) }
{\vert{\mathbf x} - {\mathbf x}_j^{PL}\vert^2}
\label{eqn:hierarchical}
\end{equation}
where $N_L$ is the number of lenses to be included directly, and
$N_{PL}$ is the number of pseudo-lenses with total mass $m_j^{PL}$ and
centre of mass position ${\mathbf x}_j^{PL}$. We note that the actual
implementation of the tree code uses higher order multipoles of the
mass distribution in order to increase accuracy. This calculation scales 
on average as $O({\rm log}N_*)$. 

Additional techniques are used in the standard hierarchical 
tree code in order to improve the accuracy and speed of the calculation. For
example, ``test rays'' are used to determine the cell structure, and
then the same cells are used for a number of actual rays surrounding
the test ray. Also, the deflection due to pseudo-lenses is only
calculated directly for the test rays; for the actual rays,
pseudo-lens deflections are interpolated between those
calculated for the test rays. For full details, see \citet{wambsganss99}.

\subsection{CPU-P}
A tree code requires the following data structures 
to be stored in memory: 2-D magnification map array; 
lens array (mass and 2-D location for each lens); and the cell tree 
(stored as an array of indices), which provides a significant memory 
overhead as $N_*$ increases.  In order to increase the scale of microlensing
simulations beyond those that were achievable with CPU-T, where
 typical single-core memory permits a maximum of ~20 million lenses, \citet{garsden+10} implemented a distributed, parallel version of the tree 
code (CPU-P).  They take advantage of the inherent parallelism of the ray 
deflections by distributing rays between multiprocessors 
of a CPU-based cluster supercomputer, using a simple but effective approach to 
static load balancing between computing nodes. To handle the large 
memory requirements for billion-lens simulations ($\sim 200$ GB), they 
use a combination of local memory, distribution of 
data between nodes (thus sharing the memory requirements), and 
disk-based storage (although this limits processing speeds somewhat by 
the requirement of costly file accesss).

The CPU-P code implements the same cell-tree and microlensing
algorithms as CPU-T, but surrounded by a quite different environment within a
running simulation.
The generation of lens masses at the beginning of the run, and the
ray-shooting itself,  use as much parallelism as is available. The
generation of the cell tree has not been parallelised, but  is  executed
concurrently with another phase in the simulation. All these
operations are logically  the same as in the CPU-T version but execute
without knowledge that they are running concurrently, or that their data
may be contained within  huge disk files. Disk files  naturally degrade performance and parallel speed-up, but
there are various caching
schemes which can make up for that.
CPU-P has been built to have the same accuracy as CPU-T, because the same numerical
algorithms are used, however due to the reimplementation
in a different programming language (FORTRAN for CPU-T, C++ for CPU-P), with different random number
generators, and a different precision
of integers (arbitrary precision has been implemented), there will
naturally be minor differences.

\subsection{GPU-D}

The GPU parallel implementation of direct ray-shooting is achieved by 
splitting the total number of light rays into $N_{\rm B}$
batches, which are deflected in parallel. For convenience, the
deflection calculation of equation (\ref{eqn:sigma}) is recast as:
\begin{equation}
\left(
\begin{array}{cc}
1 - \gamma & 0 \\
0 & 1 + \gamma
\end{array}
\right)
{\mathbf x}  + \sigma_c {\mathbf x} - {\mathbf y} 
= \sum_{t=1}^{T} \sum_{i=1}^{N_t} m_i 
\frac{({\mathbf x} - {\mathbf x}_i ) }
{\vert{\mathbf x} - {\mathbf x}_i\vert^2}
\end{equation}
where $\sum_{t=1}^{T} N_t = N_*$ for $N_{\rm T}$ parallel threads, and each
processing step considers only a single light ray.    The contributions 
from $\gamma$ and $\sigma_c$ are determined at the end of the 
deflection calculation.

The GPU-D code, described in Thompson et al. (2010), was implemented in CUDA according to the following algorithm:
\begin{enumerate}
\item $N_*$ lens coordinates are obtained (see below) and stored in CPU memory. 
Coordinates are loaded into GPU device memory as required;
\item $N_{\rm B} = 2^{17}$ light ray coordinates are pseudo-randomly generated 
on the CPU and loaded into GPU device memory;
\item GPU computation is initialised; computation is split into groups of
$N_{\rm T} = 128$ threads;
\item Each thread group loads $N_{\rm T}$ lenses and calculates 
deflection on $N_{\rm T}$ rays; this is repeated until all lenses and 
rays are exhausted; 
\item Once computation on the GPU is complete, results are copied
 back from the GPU device memory to system memory;
\item CPU maps the ray locations onto the source pixel grid in order to obtain the magnification map;
\item Steps 2-6 repeated until an average of $N_{\rm rays}$ rays per source pixel is reached.
\end{enumerate}
The only change in approach from that of Thompson et al. (2010) is the
way the $N_{*}$ lenses coordinates are generated: for this work, we 
first run a tree code and output lens positions to a data file.  
These coordinates are then read from the data file by the GPU-D 
code, and subsequent processing
is carried out as described above.  Without this modification, we would not be 
able to make pixel-by-pixel comparisons of the magnification maps, and would be
left with just statistical comparisons based on, for example, 
magnification probability distributions.
For full details on choice of optimisation of GPU parameters, especially 
$N_{\rm B}$ and $N_{\rm T}$, see \citet{thompson+10}, as these are
somewhat hardware dependent.

\section{Comparing the methods}
\label{sct:results}
In this section, we consider the relative accuracies of CPU-T, CPU-P 
and GPU-D over a much wider range of parameter space than has 
been considered previously, and we compare processing times 
between CPU-T and GPU-D, noting regions of parameter space where 
each code might currently be preferred.  The former aspect is 
important in assessing the validity of all microlensing results that 
have relied on the tree code to date,  the latter in motivating a choice
of code for future applications.

The choice of model parameters for comparison is somewhat
arbitrary. We attempt to cover the majority of $\sigma$ and $\gamma$
parameter space typically probed by realistic lens models, as well as
typical map sizes and pixel scales used in those analyses (see
for example: \citealt{wms95}; \citealt{k04}; \citealt{morgan+06};
\citealt{bate+08}; \citealt{mediavilla+09}).  

\subsection{Accuracy}
Since the tree code uses an approximation to the lens locations, in some
sense the direct approach produces a magnification map that is more 
accurate for a given configuration of lenses.  An unanswered question 
is whether the hierarchical approach maintains sufficient
accuracy as the number of calculations, $\Phi$, is increased beyond 
approximately $2\times10^{14}$. This restriction is largely historical, 
due to the impracticality of using the direct ray-shooting approach 
on a single-core CPU beyond this limit. 

We conducted simulations for a range of parameters, summarised below:
\begin{itemize}
\item $N_{\rm pix} = 1024^2$, $2048^2$ and $4096^2$;
\item $N_{\rm rays} = 100$, $200$, $500$, $1000$ and $2000$; and
\item $0.1 \leq \sigma \leq 0.9$ in increments of $\Delta \sigma =
  0.1$, with $\gamma = 0.0$ and $\sigma_c = 0.0$. This corresponds to $N_* = 65$, $163$, $313$, $558$, $987$, $1817$, $3701$,
$9343$ and $41257$.
\item $0.0 \leq \gamma \leq 1.0$ in increments of $\Delta \gamma =
  0.1$, with $\sigma = \sigma_* = 0.450$ and $\sigma_c = 0.0$. This
  corresponds to $N_* = 741$, $814$, $1093$, $1875$, $4796$, $41134$, $41119$,
  $4748$, $1787$, $956$ and $606$.
\item $0.0 \leq s \leq 0.9$ in increments of $\Delta s = 0.1$ and $s = 0.99$, where the smooth matter
  fraction $s = \sigma_c/ \sigma$, $\sigma = 0.450$ and $\gamma = 0.5$.
\end{itemize}
These parameter choices allow us to probe the range:
\begin{equation}
6.8\times10^{10} \leq \Phi \leq 1.4 \times 10^{16},
\end{equation}
where we assume that $N_{\rm FLOP} = 10$. 
We choose all lenses to have the same lens mass as this does not
influence the run-time.  The CPU-T/CPU-P accuracy
parameter was set to 0.60 for all of our simulations, as
this was found to reliably produce acceptable magnification
maps. CPU-P maps were generated using a minimum of 2 compute nodes, to ensure
that the parallelisation was not producing inaccurate results.

There is a subtle difference between the way $N_{\rm rays}$ is defined for
CPU-T/CPU-P and GPU-D. In both codes, the same number of rays are
shot across the entire shooting grid. The CPU-T/CPU-P rays are shot on a
regular grid, thus in the absence of lensing exactly $N_{\rm rays}$
are received at each CPU-T/CPU-P pixel. The GPU-D rays
are shot from random positions in the shooting grid, and so \textit{on
  average} $N_{\rm rays}$ are received at
each GPU-D pixel. This will introduce Poisson noise into the GPU-D
magnification maps, however the total number of rays shot is so large
($10^8$ to $4\times10^{10}$ in our simulations) that any error
introduced by shooting rays randomly will be insignificant.

We use two methods for assessing accuracy: the root mean square difference 
between corresponding magnification maps, and a
Kolmogorov-Smirnov (K-S) test on probability distributions for change
in magnitude $\Delta m$. The K-S test results in a measurement of the
significance with which we can reject the null hypothesis, that the
two samples were drawn from the same distribution. Change in magnitude is calculated from the
(per pixel) map magnifications using the following relationship:
\begin{equation}
\Delta m_{ij} = -2.5{\rm log}_{10}(\mu_{ij}).
\label{eqn:deltam}
\end{equation}
Throughout, we define 
``excellent agreement'' as a root mean square difference between maps 
of less than 0.2 magnitudes (often much less than 0.2 magnitudes), and 
a K-S test probability that the null hypothesis is not rejected of greater than
$99.7 \%$. 

\begin{table}
\caption{Comparison between the CPU-T, CPU-P and GPU-D codes for 
magnification maps with $N_{\rm pix} = 4096^2$, $\gamma = 0.0$ and 
$s=0.0$ (corresponding to $\sigma_c = 0.0$). $RMS$ is the root mean square difference between magnification
maps (compared pixel by pixel), and $P(KS)$ is the K-S test
probability. The CPU-T and CPU-P comparisons are virtually indistinguishable.
\label{tbl:Res1}}
\begin{tabular}{cccccc}
\hline &&
\multicolumn{2}{c}{CPU-T and GPU-D} &
\multicolumn{2}{c}{CPU-P and GPU-D} \\ 
$N_{\rm rays}$ & $\sigma$ & $RMS$ & $P(KS)$ & $RMS$ & $P(KS)$ \\
\hline
100 & 0.1 & 0.782 & 0.247 & 0.782 & 0.247 \\
& 0.2 & 0.757 & 0.389 & 0.757 & 0.389 \\
& 0.3 & 0.734 & 0.000 & 0.733 & 0.000 \\
& 0.4 & 0.712 & 0.004 & 0.712 & 0.000 \\
& 0.5 & 0.731 & 0.802 & 0.731 & 0.802 \\
& 0.6 & 0.730 & 0.277 & 0.730 & 0.277 \\
& 0.7 & 0.736 & 1.000 & 0.736 & 1.000 \\
& 0.8 & 0.799 & 1.000 & 0.799 & 1.000 \\
& 0.9 & 0.813 & 1.000 & 0.813 & 1.000 \\
200 & 0.1 & 0.390 & 0.010 & 0.390 & 0.010 \\
& 0.2 & 0.415 & 0.413 & 0.415 & 0.413 \\
& 0.3 & 0.442 & 0.473 & 0.442 & 0.473 \\
& 0.4 & 0.495 & 0.835 & 0.494 & 0.835 \\
& 0.5 & 0.490 & 1.000 & 0.490 & 1.000 \\
& 0.6 & 0.536 & 0.856 & 0.536 & 0.856 \\
& 0.7 & 0.572 & 1.000 & 0.572 & 1.000 \\
& 0.8 & 0.559 & 1.000 & 0.559 & 1.000 \\
& 0.9 & 0.601 & 1.000 & 0.601 & 1.000 \\

500 & 0.1 & 0.143 & 0.015 & 0.143 & 0.015 \\
& 0.2 & 0.150 & 0.239 & 0.150 & 0.239 \\
& 0.3 & 0.154 & 0.997 & 0.154 & 0.997 \\
& 0.4 & 0.168 & 0.999 & 0.168 & 0.999 \\
& 0.5 & 0.166 & 1.000 & 0.166 & 1.000 \\
& 0.6 & 0.185 & 1.000 & 0.185 & 1.000 \\
& 0.7 & 0.206 & 1.000 & 0.206 & 1.000 \\
& 0.8 & 0.191 & 1.000 & 0.191 & 1.000 \\
& 0.9 & 0.212 & 1.000 & 0.212 & 1.000 \\
1000 & 0.1 & 0.075 & 0.039 & 0.075 & 0.039 \\
& 0.2 & 0.071 & 0.774 & 0.071 & 0.774 \\
& 0.3 & 0.074 & 1.000 & 0.074 & 1.000 \\
& 0.4 & 0.081 & 1.000 & 0.081 & 1.000 \\
& 0.5 & 0.079 & 1.000 & 0.079 & 1.000 \\
& 0.6 & 0.087 & 1.000 & 0.087 & 1.000 \\
& 0.7 & 0.096 & 1.000 & 0.096 & 1.000 \\
& 0.8 & 0.090 & 1.000 & 0.089 & 1.000 \\
& 0.9 & 0.099 & 1.000 & 0.099 & 1.000 \\
2000 & 0.1 & 0.042 & 0.822 & 0.042 & 1.000\\
& 0.2 & 0.035 & 1.000 & 0.035 & 1.000 \\
& 0.3 & 0.037 & 1.000 & 0.037 & 1.000 \\
& 0.4 & 0.040 & 1.000 & 0.039 & 1.000 \\
& 0.5 & 0.039 & 1.000 & 0.039 & 1.000 \\
& 0.6 & 0.043 & 1.000 & 0.043 & 1.000 \\
& 0.7 & 0.047 & 1.000 & 0.047 & 1.000 \\
& 0.8 & 0.044 & 1.000 & 0.044 & 1.000 \\
& 0.9 & 0.050 & 1.000 & 0.050 & 1.000 \\
\hline
\end{tabular}
\end{table}

\subsubsection{No shear}
To begin with, we set both the external shear, $\gamma$, and the smooth matter 
fraction, $s = \sigma_c/\sigma$, to zero.  This allows us to consider
just the effect of compact objects, which are directly related to the number of 
lenses via equation (\ref{eqn:nstar}). Later, we allow both $\gamma$ and $s$ 
to vary.

As indicative results, we present the calculated RMS values and K-S 
probabilities, $P(KS)$, 
for the $N_{\rm pix} = 4096^2$ case in Table \ref{tbl:Res1}.
For $\sigma \geq 0.3$, we find excellent agreement between the 
magnification maps from CPU-T and GPU-D when $N_{\rm rays} \geq 500$.
The $\sigma = 0.1$ and 0.2 cases differ slightly. Here, a low number
of lenses ($N_*= 65$ and $163$ respectively)
mean that there is very little variation in magnification across the
maps. The K-S test probabilities are therefore strongly affected by
small differences in magnification probability distributions. For $N_{\rm pix} = 1024^2$ and $2048^2$, $N_{\rm
  rays}=1000$ are sufficient to obtain agreement between CPU-T and GPU-D. For $N_{\rm pix} = 4096^2$ even $N_{\rm
  rays} = 2000$ are not enough. In other words, a larger number of
rays per pixel are required for agreement between CPU-T and GPU-D when
$\sigma$ is low ($\leq 0.2$).

We note also that for a given $\sigma$ and $N_{\rm rays}$, the RMS
error tends to increase with increasing $N_{\rm pix}$. For example,
when $\sigma=0.4$ and $N_{\rm rays}=500$, we obtain the following RMS
differences: 0.039 mag ($N_{\rm pix}=1024^2$), 
0.081 mag ($N_{\rm pix}=2048^2$)
and 0.168 mag ($N_{\rm pix}=4096^2$). In other words, for a fixed accuracy
we must increase $N_{\rm rays}$ by a factor of $\sim2$ 
if we increase $N_{\rm pix}$ by a factor of 4. 

For the comparisons between CPU-P and GPU-D, we find the same
results. In fact, the CPU-T and CPU-P maps are virtually
indistinguishable down to the accuracy of our comparison. They
therefore behave identically when compared with GPU-D.

\subsubsection{Shear and convergence}
We then fixed the convergence at $\sigma=0.45$, and examined
the impact of varying shear over the range $0.0 \leq \gamma \leq 1.0$ in 
increments of $\Delta \gamma = 0.1$. The number of pixels was fixed at
$N_{\rm pix} = 2048^2$. Once again, the smooth matter
fraction $s$ was set to zero. We find very similar results to the no-shear case discussed above: for $N_{\rm rays} \geq 500$, 
the results of CPU-T and GPU-D are indistinguishable (RMS difference 
$< 0.2$ magnitudes and K-S probability $> 99.7\%$), and similarly for
comparisons between CPU-P and GPU-D.

\subsubsection{Smooth matter fraction}
\label{sub:smooth_acc}
Finally, we test the accuracy of CPU-T and CPU-P when the convergence 
is split
into a continuously distributed component $\sigma_c$ and a compact
stellar component $\sigma_*$. We set $\gamma=0.5$ and $\sigma =
\sigma_c + \sigma_* = 0.45$, and vary $s$ in the range $0.0 \leq s \leq
0.9$ in increments of $\Delta s = 0.1$.  Since there is no microlensing if 
$s=1$, we also simulated the $s=0.99$ case. For convenience, we choose 
$N_{\rm pix} = 2048^2$.

For $s \leq 0.8$, we find the same results as the two cases discussed
above: for $N_{\rm rays} \geq 500$ the results of CPU-T and GPU-D are
in excellent agreement. The $s = 0.9$ case has a slightly higher 
RMS difference at $N_{\rm rays} = 500$ of $0.4$, although the 
K-S probability is still $>99.7\%$. For the $s=0.99$ case, however, 
no agreement is reached between CPU-T and GPU-D. Visual
inspection of the maps shows that too few rays have been
shot in the CPU-T case. This suggests that more rays should be shot in
CPU-T for higher smooth matter fractions; $N_{\rm rays}=1000$ should
be sufficient for $s=0.90$, however more than $N_{\rm rays} > 2000$ is
required in the $s=0.99$ case. Once again, CPU-P performs identically to CPU-T. 

By comparing the accuracy of the two tree codes, which rely on a series 
of approximations to the lens distribution in order to speed up 
calculation time, with GPU-D, which conducts the entire brute-force 
calculation, we have verified the accuracy of the standard tool for 
high optical depth microlensing simulations. Indeed, we have considered
a wider range of parameters, as represented through the quantity $\Phi$, 
than previous work ($\Phi>2\times10^{14}$). While not unexpected, our results 
confirm that the tree code can indeed be used with confidence, provided 
a sufficiently large number of rays per pixel are shot (i.e. typically
$N_{\rm rays} \geq
1000$ for the parameter space explored here). 

\subsection{Timing Tests}

\begin{figure*}
  \centering
  \scalebox{0.8}{\includegraphics[angle=270]{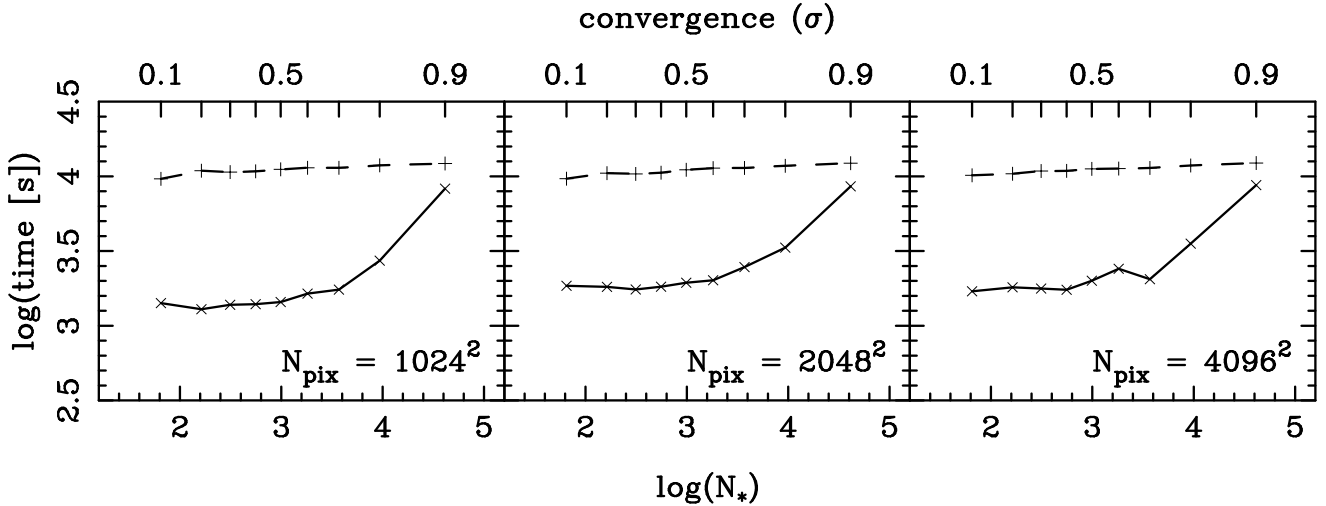}}
  \caption{Timing results for GPU-D (solid line) and CPU-T (dashed
    line) as a function of convergence, $\sigma$, or equivalently,
    number of lenses, $N_*$. In all
    cases, shear $\gamma = 0.0$, $N_{\rm rays} = 1000$ and $s=0$
    (i.e. all of the convergence is in a compact stellar component,
    $\sigma=\sigma_*$). Panels are $N_{\rm pix} = 1024^2$ (left);
    $N_{\rm pix} = 2048^2$ (middle); $N_{\rm pix} = 4096^2$ (right).
  }
  \label{fig:convergence}
\end{figure*}

Having established accuracy of the three codes under consideration,
we now turn to timing tests.  Specifically, we determine the time 
taken to generate magnification maps with a given set of model parameters, 
allowing us to determine the regions of $(\gamma,\sigma,
N_{\rm pix})$ parameter space where one of the implementations may be 
favoured.  

We do not consider timing comparisons between CPU-T and CPU-P, as these
are  reported in Garsden \& Lewis (2010).  Since CPU-P run-times 
scale with the number of nodes available, it is difficult to establish
which configuration provides the most appropriate comparison between
CPU-P and GPU-D. Using the hardware cost as a factor in choosing between software 
alternatives, we suggest that the cost of purchasing a suitable high-end
graphics card plus computer is much lower than the set-up costs 
for a modest CPU-based computing cluster, once networking overheads and 
cluster management are considered.  We reflect further on the 
implications this has for the future of microlensing computations 
in Section \ref{sct:discussion}.

Care must be taken when quoting absolute speeds for different hardware 
-- we use our results as a guide to the relative processing speeds of
CPU-T and GPU-D, but note that alternative hardware can 
change timing by factors of a few.  
Our benchmarks were conducted on the following equipment:
\begin{itemize}
\item {\bf CPU-T}
Intel Xeon Dual Core 3060 CPU (2.4 GHz) with 1 GB RAM.
\item {\bf GPU-D} 
NVIDIA S1070 Tesla unit connected via dual PCIe x8, or higher, buses
to an Intel Q6600 Quad Core CPU (2.4 GHz) with GB RAM.  The Tesla 
S1070 comprises four GPUs, each with 240 stream processors, running at 
1.296 GHz.  Each GPU is able to access 4 GB of RAM.  The quoted peak 
processing rate is 2.488 TFLOP/s, however, for GPU-D, a peak of 1.28 TFLOP/s 
was obtained using all four GPUs (Thompson et al. 2010).  We note that 
the majority of our timing benchmarks in the present work were conducted
using only two of the four GPUs available on the Tesla S1070\footnote{This was 
a logistical issue, as there were other users who required access to the
Tesla unit while our timing tests were underway.}.
\end{itemize}

Since we have determined that the magnification maps produced by CPU-T are
accurate for $N_{\rm rays} \geq 1000$ across the majority of
parameter space, we conduct our timing tests by
comparing the total time taken by CPU-T and GPU-D to generate identical maps
with $N_{\rm rays} = 1000$, for each combination of parameters.
The results of these tests are discussed below.

\subsubsection{No shear}
Timing results for $N_{\rm pix} = 1024^2$, $2048^2$ and $4096^2$, 
with $\gamma =0$ and $s=0$, are plotted in Figure \ref{fig:convergence}. 
Each panel shows both $\sigma$ (along the top axis), and
the logarithm of $N_*$ (along the bottom axis). Note that as $s=0$ for
these simulations, $\sigma = \sigma_*$.

The performance of CPU-T is
essentially constant as $\sigma$ increases. The same is not true
for GPU-D; it performs better for lower $\sigma$
(lower $N_*$) values. This result is as expected; as
the number of lenses is increased, the direct code is
required to perform correspondingly more calculations, whereas the 
approximations in the tree code (specifically the depth to which the
cell tree is accessed) mean it performs roughly the same number 
of calculations at each step.

The shape of the GPU-D timing curves is the same as those presented in
Figure 2 of \citet{thompson+10}. For low numbers of lenses ($N_*
\lesssim 10^4$), overheads induced by the need to transfer data
between GPU and CPU affect the performance of GPU-D. For $N_* \gtrsim
10^4$, near the limit of our comparison, the performance of the code
scales linearly with $N_*$. While this linear growth suggests that 
there is a cross-over in run-time between CPU-T and GPU-D at 
around $N_* = 10^5$, it must be recalled that this is approaching 
the maximum lens limit of CPU-T, due to the memory requirements 
for storing the cell tree.  

We note that although the performance of GPU-D is decreasing as
$\sigma$ (or equivalently, $N_*$) is increased, it is still faster
than CPU-T for all of our $\gamma=0.0$ simulations.  Using all four GPUs on the Tesla
S1070 unit, instead of the two we use for this detailed 
timing comparison, results in factor of two speed-up.

\subsubsection{Shear and convergence}
Real lensing galaxies are unlikely to be perfectly spherically symmetric, and so
realistic lens models do not contain zero external shear. We therefore also
conducted timing tests for varying $\gamma$. Equation \ref{eqn:sigma}
shows that the actual calculation of the effect of external shear
should not significantly add to computation time -- it simply adds one
extra operation to the calculation of the deflection of each ray. However,
the side lengths $L_{\pm}$ of the rectangular shooting grid that is projected on
to a square magnification map with side length $L_{\rm map}$ depends on the shear in the following
way:

\begin{equation}
L_{\pm} = \frac{L_{\rm map}}{1 - \sigma \pm \gamma}
\label{eqn:square}
\end{equation}
When $1-\sigma-\gamma\approx 0$ or $1-\sigma+\gamma \approx 0$, one side of the shooting grid can be
very large. This leads to a large area for the lens plane, and hence a
correspondingly large number of lenses (see Equation \ref{eqn:nstar}
and the discussion following it).

\begin{figure}
  \centering
  \scalebox{0.4}{\includegraphics[angle=270]{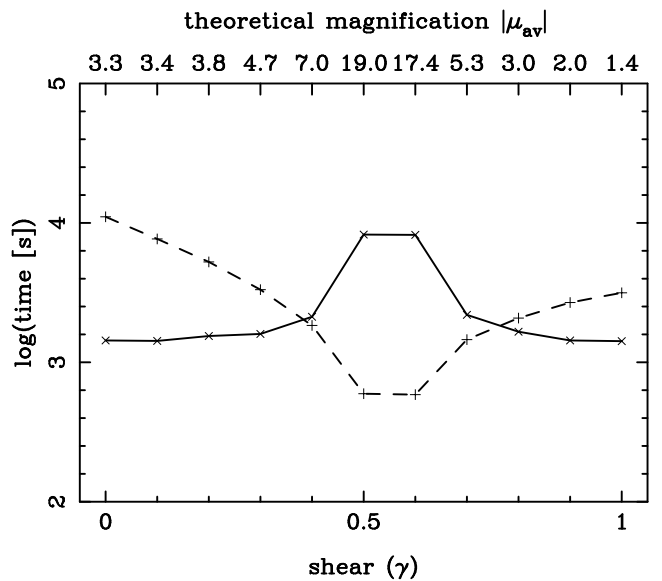}}
  \caption{Timing results for GPU-D (solid line) and
    CPU-T (dashed line) as a function of shear
    $\gamma$. The number
    of pixels $N_{\rm pix} = 2048^2$, smooth matter fraction $s=0$, and the convergence is
    $\sigma=\sigma_*=0.45$. An average of $N_{\rm rays} =1000$ was shot per
    pixel. The top axis shows the theoretically predicated average
    magnification in the absence of microlensing.} 
\label{fig:gamma}
\end{figure}

We conducted timing tests for the case where $N_{\rm rays}=1000$,
$N_{\rm pix} = 2048^2$, $s=0$ and
$\sigma = \sigma_* = 0.45$, with $0 \leq \gamma \leq 1.0$ and $\Delta \gamma = 0.1$. 
The results of these tests are presented in Figure
\ref{fig:gamma}. The figure shows $\gamma$ versus the
logarithm of the time
taken to generate the map. We also plot time against the
theoretically predicted average magnification $\mu_{\rm av}$ in that
figure, where $\mu_{\rm av}$ is given by:
\begin{equation}
\mu_{\rm av} = \frac{1}{(1 - \sigma)^2 - \gamma^2}.
\label{eqn:mutheory}
\end{equation}
A negative magnification is interpreted as a parity flip in the
image. 

For low magnifications (when $1-\sigma-\gamma$ and $1-\sigma+\gamma$ are very
different from zero), GPU-D is faster than CPU-T. There
is a crossover point at a magnification of $\sim5$ where CPU-T 
begins to out-perform GPU-D. For high
magnifications, CPU-T is
significantly faster than GPU-D. We emphasise again that these
timing tests were conducted using only two of the available four
GPUs. Running on all four GPUs should improve the performance of GPU-D
by approximately a factor of two.

The timing behaviour of GPU-D with the addition of a non-zero shear is
exactly the same as the zero shear case. Initially, overheads dominate
the code's performance, but as the number of lenses is
increased to $\sim10^4$, the code begins to scale linearly with
$N_*$. 

The same is not true for CPU-T -- the timing behaviour in the
zero shear and varying shear cases seems to be quite different. This
is due to the way in which the shooting grid is stretched by the
addition of a non-zero shear. Equation \ref{eqn:square} shows that for
certain combinations of $\sigma$ and $\gamma$ the shooting grid will be
significantly stretched in one direction. A ray shot from one corner
of this stretched grid is close to fewer lenses than a ray shot from a
corner of a square with the same area and thus a greater
number of lenses can be grouped together into pseudo-lenses. Although
$N_*$ goes up for these cases, the number of lenses that need to be included directly
in the calculation goes \textit{down} for the tree code.

We expect this behaviour to be generic as $\sigma$ is varied -- CPU-T
will be faster than GPU-D in the high
magnification regime on current-generation hardware, whereas GPU-D will out-perform CPU-T in the low magnification
regime.  However, as we discuss in Section \ref{sct:discussion}, the former
situation may only be a limit in the short-term.

To provide a physical context for these results, the \citet{k04} model
for image $A$ in the lensed quasar \astrobj{Q2237+0305} has $\sigma =
0.394$ and $\gamma = 0.395$, and a corresponding theoretical
magnification $\mu_{\rm av} = 4.735$. This is in line with most models
for \astrobj{Q2237+0305} image A, which typically have $\sigma \sim
0.4$, $\gamma \sim 0.4$ and $\mu_{\rm av} \sim 5$ (e.g. Wyithe,
Agol \& Fluke 2002). In comparison, the
\citet{keeton+06} model for the high magnification image $D$ in \astrobj{SDSS J0924+0219} has $\sigma = 0.476$, $\gamma = 0.565$ and a magnification of
$\mu_{\rm av} = -22.4$.

\subsubsection{Smooth matter fraction}

\begin{figure}
  \centering
  \scalebox{0.4}{\includegraphics[angle=270]{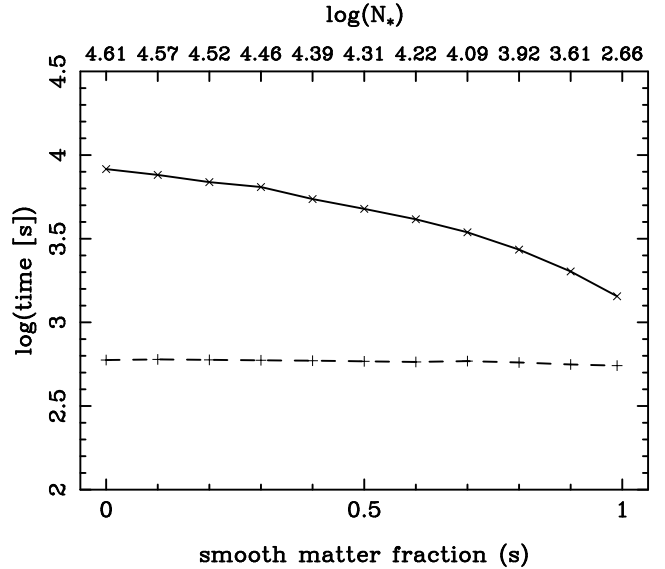}}
  \caption{Timing results for GPU-D (solid line) and
    CPU-T (dashed line) as a function of smooth matter fraction
    $s$. The total convergence $\sigma=0.40$, shear $\gamma=0.5$, the number
    of pixels $N_{\rm pix} = 2048^2$, and the average number of rays
    per pixel is    $N_{\rm rays} = 1000$.} 
\label{fig:smooth}
\end{figure}

Finally, we present the results of the timing tests for varying smooth
matter fraction in Figure \ref{fig:smooth}. Tests were
conducted for $N_{\rm pix} = 2048^2$, $\sigma = 0.45$, $\gamma = 0.5$,
and $N_{\rm rays} = 1000$. This combination of $\sigma$ and $\gamma$
corresponds to the ``worst case'' magnification for GPU-D timing, and
the ``best case'' magnification for CPU-T timing (see the previous section).

Increasing the smooth matter fraction for fixed $\sigma$ corresponds
to decreasing $N_*$, as mass is transferred from the compact
stellar component to the continuously distributed component. We therefore see exactly the timing behaviour we
expect: CPU-T takes roughly the same amount of time for each $s$,
whereas GPU-D is faster for fewer lenses. 

At the distances from the centre of lensing galaxies
where lensed images are typically observed ($\sim2-10$ kpc), we expect
the smooth matter fraction to be quite high. Indeed, a few rough
microlensing measurements of smooth matter percentage at lensed image
positions exist: 80\% to 85\% at the location of the $A$ and $D$
images in \astrobj{SDSS J0924+0219} \citep{keeton+06}; $\sim80\%$ in \astrobj{HE 1104+080}
\citep{chartas+09}; $\sim90\%$ in \astrobj{PG 1115+080} \citep{pooley+09};
and $\sim70\%$ in \astrobj{RXJ 1131-1231} \citep{dai+10}. For smooth matter
fractions in this range, we find that GPU-D already performs
comparably to CPU-T in high magnification systems, and out-performs it
in low magnification systems.

We note that our accuracy tests in Section \ref{sub:smooth_acc} showed
that even $N_{\rm rays} = 2000$ were not sufficient for CPU-T to
obtain results indistinguishable from GPU-D at $s=0.99$, at least with
the CPU-T accuracy parameter used throughout this paper. The $s=0.99$
point in the Figure \ref{fig:smooth} CPU-T results is therefore likely
to be lower than the actual run-time required to produce a useable map.

\section{Discussion}
\label{sct:discussion}
Our intention in this work is not to imply that the tree code is 
excessively inaccurate. However, its reliance on 
an approximation to the lens distribution to reduce the number of 
calculations per magnification map warrants further investigation that 
has not been feasible until now. 
The standard tree code (Wambsganss 1990; 1999) has been in widespread 
use throughout the lensing community since the early 1990s. 
Though it is unsurprising, it is nevertheless reassuring to know that
CPU-T remains accurate over an extended range of $\Phi$-space. 
Our comparison here demonstrates that researchers 
can, and should, continue to use the tree code for microlensing simulations 
with confidence.  Additionally, we have demonstrated that this accuracy is
also maintained by the CPU-P code, which enables exploration of problems
orders of magnitude beyond those that can be achieved on a single-core CPU.

However, we stress that the direct inverse ray-shooting code {\em is} 
more accurate than its tree code counterparts by simple virtue that
it includes contributions from all lenses without 
approximations -- the entire calculation is done by 
brute force, taking advantage of the massively parallel GPU architecture 
to complete the computation in a reasonable timeframe. 
While CPU-T can indeed be used for a wide range of astrophysically relevant 
scenarios, GPU-D is now a highly competitive alternative.  

Given that we have verified the accuracy of
the tree codes, though, it is reasonable to ask in which areas of
parameter space each code should be used. The larger the number of
lenses, the better the tree codes perform relative to GPU-D.
For the values of $N_{\rm pix}$ and $N_{\rm rays}$ we examined
in our benchmarking, the crossover point where CPU-T begins to
out-perform its GPU-based counterpart is $N_* \sim 50,000$
lenses. This can occur either when $(\sigma-\gamma)$ or $(\sigma+\gamma)$
are close to 1.  For scenarios with 
$N_{\rm pix} > 4096^2$ and $N_* > 10^5$, there are advantages in 
moving to a distributed CPU-based computing cluster, although this can be 
a financially costly approach.  One of the perceived benefits of 
GPU computing is the heavily reduced dollar per Tflop/s they offer, 
which makes them an attractive alternative as ``desktop'' supercomputers
(see \citealt{thompson+10}). The downside of GPUs, at this point in time, 
is the paucity of astronomy codes that have been adapted to run on them,
whereas CPU-based computing clusters can support a much wider range of existing 
codes. 

Since about 2005, single-core CPU processing speeds have plateaued.
Additional computing performance has been achieved, and indeed Moore's Law
growth in processing power has been maintained, by moving to multi-core
CPUs.  A growing range of applications across a variety of scientific 
disciplines are now taking advantage of the ``many-core'' processing 
architecture of GPUs, which are providing a preview of the future of CPU
architectures.  Yet if GPUs do indeed point towards a future where 
many-core processing is mainstream, 
the need to convert existing trusted, tested, accepted codes will 
become more pressing.  A question that many researchers will have to 
answer is which code should be adapted? 

We suggest that decisions to move other standard astronomy computations to 
GPU may benefit from a similar approach to Thompson et al. (2010). Rather
than attempting to directly port an existing advanced code (i.e. CPU-T) to
GPU, they considered a brute force approach that was not feasible for
single-core CPU, but which exhibited the type of fine-grained, massive 
parallelism that is ideal for GPGPU. This resulted in much faster code
development time, despite the learning curve associated with adopting
CUDA and the specifics of GPU parallel programming.
Moreover, we have shown that in most astrophysically-motivated cases, 
we are already {\em no worse off} using GPU-D compared with CPU-T, 
with the added benefit of greater intrinsic accuracy.

The $N_*$-crossover in timing tests that we have identified 
does depend on the hardwared used.  For example, our benchmarking 
was performed using two of the four GPUs available on the Tesla S1070,
while the CPUs used for CPU-T were a few years old. The key point, 
however, is that the GPU-D is already parallel -- GPU performance 
will increase, and more GPU cores will be placed in individual machines, yet
no further code development is required. 
The single-core tree code, however, is more heavily restricted -- 
CPU-T can no longer make significant processing gains by simply 
relying on Moore's Law to make single-core CPU hardware faster. 

\section{Conclusion}
\label{sct:conclusion}
In this work, we have examined the accuracy of three highly competive 
alternatives for generating microlensing magnification maps: 
a widely-used hierachical tree code for single-core CPUs; a parallel, large
data implementation for a distributed CPU-based computing cluster; and a brute force
solution that utilises the massive parallelism of modern graphics processing 
units.  
We have demonstrated that these three codes do indeed present 
comparable accuracy, and choice of approach will depend on considerations
relating to the scale and nature of the problem to be investigated.

While considered impractical as a calculating approach even a few years ago, 
our timing tests show that direct, inverse ray-shooting is now a viable, 
and indeed more accurate, option.  Moving a simple
algorithm to GPU, rather than attempting to directly implement a more
advanced solution -- originally developed to overcome a hardware restriction
of single-core CPUs -- to achieve run-times that are already similar,
suggests an approach that other early adopters
of GPUs for astronomy computation may wish to consider.


\section*{Acknowledgements}
We wish to thank Alex Thompson for the original GPU-D code development.  We are extremely 
grateful to Joachim Wambsganss for providing his microlensing tree code, 
which continues to be of great value to the gravitational microlensing 
community.  CJF thanks David Barnes for valuable discussions relating to 
the philosophy of GPU programming. This research was supported under 
the Australian Research Council's Discovery Projects funding scheme 
(project number DP0665574).

\bibliographystyle{model2-names}



\end{document}